\documentclass[letterpaper,twoside,twocolumn,english,prl,intlimits,showpacs]{revtex4}
\usepackage{times}
\usepackage[T1]{fontenc}
\usepackage[latin2]{inputenc}
\usepackage{amsmath}
\usepackage{graphicx}
\usepackage{amssymb}

\makeatletter

\usepackage{babel}
\makeatother
\begin{document}

\title{Particle Dispersion on Rapidly Folding Random Hetero-Polymers}

\author{D.~Brockmann}

\author{T.~Geisel}

\affiliation{Max-Planck-Institut für Strömungsforschung and Fakultät für Physik,
Universität Göttingen, 37073 Göttingen, Germany}

\begin{abstract}
We investigate the dynamics of a particle moving randomly along a
disordered hetero-polymer subjected to rapid conformational changes
which induce superdiffusive motion in chemical coordinates. We study
the antagonistic interplay between the enhanced diffusion and the
quenched disorder. The dispersion speed exhibits universal behavior
independent of the folding statistics. On the other hand it is strongly
affected by the structure of the disordered potential. The results
may serve as a reference point for a number of translocation phenomena
observed in biological cells, such as protein dynamics on DNA strands.
\end{abstract}

\pacs{05.40.-a, 82.35.-x, 02.50.-r, 45.10.Hj}

\maketitle

\newcommand{\differential}[1]{\textrm{d}#1}

\newcommand{\diff}[1]{\differential{#1}}

\newcommand{\dx}{\differential{x}}

\newcommand{\dy}{\diff{y}}

\newcommand{\pd}[1]{\partial_{#1}\, }

\newcommand{\pdt}{\pd{t}}

\newcommand{\dk}{\differential{k}}

\newcommand{\fraclp}[2]{\Delta_{#2}^{#1}}

\newcommand{\fraclpmu}[1]{\fraclp{\mu/2}{#1}}

\newcommand{\laplace}[1]{\fraclp{}{#1}}

\newcommand{\op}[1]{\mathcal{#1}}

\newcommand{\zvec}[1]{\mathbf{#1}}

The study of random motion on complex structures is essential to the
understanding of dispersion phenomena observed in numerous physical
systems, ranging from epidemics spreading in complex networks and
information transport in modern communication networks such as the
internet~\cite{reka_00047:2002,volche_046137:2002}. In biological
cells, the transport of macromolecules is accomplished by a variety
of translocation processes in which carrier molecules move along complex
fibrous polymer networks, e.g. myosin translocation on actin fibers~\cite{ambla_04470:1996}
or transport on microtubules~\cite{caspi_05655:2000}. If the involved
topologies are scale-free, diffusion is often anomalous, i.e. the
mean square displacement of a particle violates the linear dependence
on time $\left\langle X^{2}(t)\right\rangle \sim t^{\gamma}$ with
$0<\gamma\ne1$~\cite{bouch_00127:1990}. Depending on the underlying
microscopic dynamics, subdiffusive ($\gamma<1$) as well as superdiffusive
($\gamma>1$) behavior is observed. For instance, when a particle
moves along a polymer in a complex folding state, it may jump to a
neighboring location in Euclidean space which is distant in chemical
coordinates. Effectively, the particle moves superdiffusively along
the chain~\cite{sokol_00857:1997,manna_04337:1989} and performs
a random walk known as a Lévy flight. This mechanism may explain fast
target localization of regulatory proteins moving along DNA strands~\cite{berg_06929:1981}.
Lévy flights have been observed in a variety of systems, ranging from
chaotic systems~\cite{geise_00616:1985} and particle dispersion
in turbulent flows~\cite{porta_01017:2001} to foraging animals~\cite{viswa_00413:1996,levan_00237:1997}
and climate changes~\cite{ditle_01441:1999}. Lévy flights are characterized
by an exponent $0<\mu<2$ which quantifies the degree of superdiffusion
and is related to the heuristic dispersion relation $X(t)\sim t^{1/\mu}$.
When Lévy flights successfully mimic single trajectories, the associated
stochastic evolution equations are no longer of the Fokker-Planck
type but rather generalizations thereof which involve fractional differential
operators. Fractional models have contributed considerably to the
understanding of these systems, in fact the terms fractional kinetics
and fractional dynamics have been coined to classify them~\cite{foged_01657:1994,hilfe:2000,metze_00001:2000,brock_00409:2002,brock_170601:2003,zasla_00461:2002}.
Of particular interest are systems in which the cause for superdiffusive
dispersion and the heterogeneity of the environment interact antagonistically. 

In this Letter we introduce and investigate a model for superdiffusive
particle dispersion on flexibly folding random hetero-polymers. We
focus on the interplay between long range Lévy type transitions due
to folding and the quenched random disorder caused by the heterogeneity
of monomers of the chain. Based on simple assumptions on the hopping
rate and configurational dynamics, we derive a fractional Fokker-Planck
equation (FFPE) describing the motion of the particle along the polymer.
We compute the relaxation properties as a function of the effective
potential strength and the Lévy exponent $\mu$. We find that the
dispersion speed depends considerably on $\mu$, but becomes universal
on larger spatial scales apart from a discontinuous change at $\mu=2$
(i.e. for ordinary diffusion). Furthermore, the relative concentration
of monomers and thus the particular shape of the potential does not
affect the ordinary diffusion process ($\mu=2$), but strongly affects
all superdiffusive processes, a result we believe to be crucial for
the understanding of transport phenomena in living cells.

\begin{figure}
\begin{center}\includegraphics[%
  width=1.0\columnwidth]{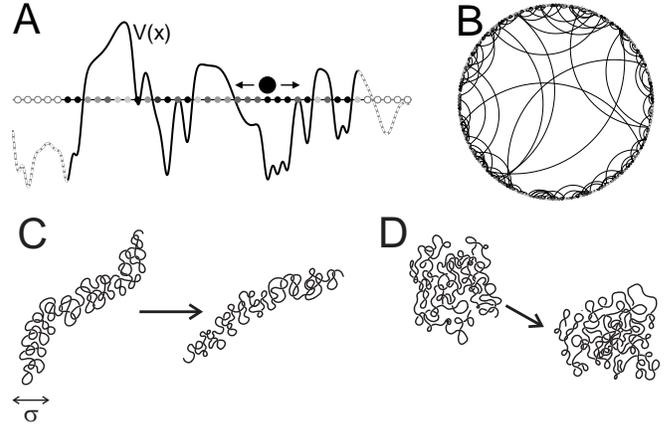}\end{center}

\caption{\label{cap:wurst1}Random hopping along heterogeneous polymers. \textbf{A:}
A particle (black disc) moving along the chemical axis $x$ experiences
a random potential $V(x)$ associated with the random sequence of
different types of monomers. When the chain is in a complex folding
state, locations that are distant along the chemical axis $x$ may
be close in Euclidean space (\textbf{C},\textbf{D}). The folding topology
is determined by a connectivity matrix depicted in \textbf{B} where
circular arcs indicate neighborhood in Euclidean space. \textbf{C:}
A folding state with a characteristic mesoscopic scale $\sigma$.
\textbf{D:} A freely flexible chain with long range connections on
all scales. Arrows indicate conformational change over time.}
\end{figure}

Consider the scenario depicted in Fig.~\ref{cap:wurst1}. A particle
is attached to a hetero-polymer and performs a random walk along the
chain. Let $x$ denote the chemical coordinate with an inter-monomer
spacing of $a$. The chain is flexible, and rapidly changing its conformational
state defined by the Euclidean coordinate $\zvec{R}(x)$ of each monomer.
The heterogeneity of the chain is modeled by a potential $V(x)$ which
specifies the probability of the particle being attached to site $x$.
In a thermally equilibrated system this probability is proportional
to the Boltzmann factor $\exp\left[-\beta V(x)\right]$. The dynamics
of the particle is governed by the rate $w(x|y,t)$ of making a transition
$y\rightarrow x$ at time $t$. We assume that transitions occur only
between monomer sites which are close in Euclidean space, i.e. when
$|\zvec{R}(y,t)-\zvec{R}(x,t)|\lesssim a$. We make the simplest possible
ansatz for this rate to take into account the requirements of Gibbs-Boltzmann
statistics, the potential heterogeneity of the chain and the complexity
of conformational states,\begin{equation}
w(x|y,t)=\frac{1}{\tau}\, e^{-\beta[V(x)-V(y)]/2}\Gamma(x,y;t).\label{eq:rate}\end{equation}
where the parameter $\tau$ is the typical microscopic time constant
and the function $\Gamma(x,y;t)$ is defined by\begin{equation}
\Gamma(x,y;t)=\begin{cases}
1 & \quad\text{if}\quad\left|\zvec{R}(x,t)-\zvec{R}(y,t)\right|\leqslant a,\\
0 & \quad\text{otherwise}.\end{cases}\label{eq:Gxyt}\end{equation}
$\Gamma(x,y;t)$ reflects the dependence of transitions on the time
dependent conformational state of the chain and is symmetric, i.e.
$\Gamma(x,y;t)=\Gamma(y,x;t)$. The function $\Gamma(x,y;t)$ can
be interpreted as a time dependent connectivity matrix (Fig.~\ref{cap:wurst1}B).
The propagator $p(x,t)$ of a particle initially ($t=0$) at the origin
evolves according to the master equation,\begin{equation}
\pdt p(x,t)=\int\dy\,\left[w(x|y,t)\, p(y,t)-w(y|x,t)\, p(x,t)\right]\label{eq:master1}\end{equation}
in which the rate is given by Eq.~(\ref{eq:rate}). The geometrical
factor $\Gamma(x,y;t)$ varies erratically and can be regarded as
a stochastic process evolving on a time scale $\tau_{g}$, which is
generally different from the hopping time constant $\tau$. Averaging
(denoted by $\left[\cdot\right]$) over conformational states the
dynamics reads $\pdt\left[p\right]=\left[\op{L}\, p\right]$ where
the operator $\op{L}$ is defined by the rhs of Eq.~(\ref{eq:master1}).
If conformational changes occur on smaller time scales than the hopping
($\tau_{g}\ll\tau$) we may substitute$\left[\op{L}\, p\right]\approx\left[\op{L}\right]\left[p\right]$,
which represents a mean field approximation. In mean field, Eq.~(\ref{eq:rate})
is given by\begin{equation}
\left[w(x|y,t)\right]=\frac{1}{\tau}e^{-\beta\left[V(x)-V(y)\right]/2}\rho(|x-y|),\label{eq:mfrate}\end{equation}
where $\rho(|x-y|)=\left[|\zvec{R}(x,t)-\zvec{R}(y,t)|\leq a\right]$
is the probability that two given sites $x$ and $y$ are neighbors
in Euclidean space. If the folding process is stationary, this probability
is time independent, and due to translation invariance along the chain
it is a decreasing function of distance in chemical space. The specific
functional form of $\rho(x)$ determines the asymptotics of Eq.~(\ref{eq:master1}).
Consider the situation depicted in Fig.~\ref{cap:wurst1}C., where
the chain is knotted such that non-local transitions occur on a typical
scale $\sigma>a$. On larger scales $\rho(x)$ vanishes. In this case,
a Kramers-Moyal expansion of the rhs of Eq.~(\ref{eq:master1}) yields
the FPE $\pdt p=\nabla V^{\prime}p+D\laplace{}p$, in which the diffusion
coefficient is given by $D\sim(\sigma/a)^{2}/\tau$ and the gradient
force is determined by the potential $V(x)$ along the chain. The
situation changes drastically for the type of chain sketched in Fig.~\ref{cap:wurst1}D.
For a freely flexible chain the quantity $\rho(x)$ follows an inverse
power law with increasing chemical distance, i.e. $\rho(x)\sim1/|x|^{1+\mu}$.
Typically $\mu<2$~\cite{degennes:1979} and thus $\rho(x)$ lacks
a well defined variance and consequently a typical scale in long range
transitions. A particle moving along such a chain will behave superdiffusively
and perform a Lévy flight in chemical coordinates. Inserting $\rho(x)\sim1/|x|^{1+\mu}$
with $0<\mu<2$ into Eq.~(\ref{eq:mfrate}) and subsequently into
Eq.~(\ref{eq:master1}) the asymptotics is governed by a fractional
Fokker-Planck equation (FFPE),\begin{equation}
D^{-1}\pdt p=e^{-\beta V/2}\,\fraclpmu{}\, e^{\beta V/2}\, p-p\, e^{\beta V/2}\,\fraclpmu{}\, e^{-\beta V/2}.\label{eq:FFPE}\end{equation}
A detailed derivation is given in Ref.~\cite{brock_00409:2002}.
Here, $D$ is the generalized diffusion coefficient and the operator
$\fraclpmu{}$ is a generalization of the ordinary Laplacian, \begin{equation}
\left(\fraclpmu{}f\right)(x)=C_{\mu}\int\diff{y}\,\frac{f(y)-f(x)}{|x-y|^{1+\mu}}\label{eq:fraclap}\end{equation}
with $C_{\mu}=\pi^{-1}\Gamma(1+\mu)\sin(\pi\mu/2)$. In contrast to
the ordinary Laplacian, $\fraclpmu{}$ is a non-local, singular integral
operator, reflecting the superdiffusive behavior of the process. The
boundary case $\mu=2$ represents the limit of ordinary diffusion,
i.e. Eq.~(\ref{eq:FFPE}) reduces to an ordinary FPE. When the potential
vanishes, $V\equiv0$, Eq.~(\ref{eq:FFPE}) becomes $\pdt p=D\fraclpmu{}p$
and is solved by the propagator of the symmetric Lévy stable process
of index $\mu$, i.e. $p(x,t)=(Dt)^{1/\mu}\, L_{\mu}\left(x/(Dt)^{1/\mu}\right)$
with $L_{\mu}(z)=(2\pi)^{-1}\int\dk\,\exp(ikz-|k|^{\mu})$~\cite{brock_00409:2002}.
In its general form Eq.~(\ref{eq:FFPE}) describes the dynamics of
Lévy flights in external potentials obeying Gibbs-Boltzmann thermodynamics.

\begin{figure}
\includegraphics[%
  width=1.0\columnwidth]{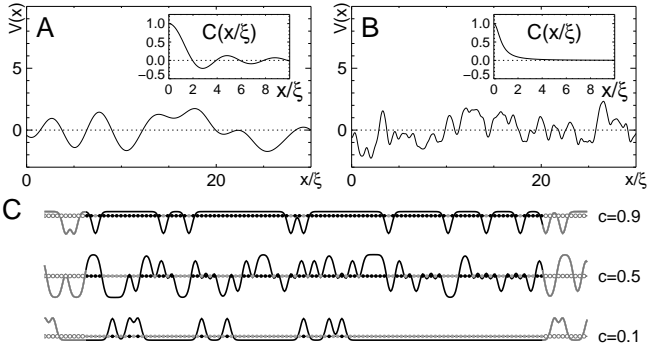}

\caption{\label{cap:fig2}\textbf{A:} A random phase potential with a power
spectrum $S(k)=2\xi\,\Theta(|k-\pi/2\xi|)$ where $\Theta$ is the
Heaviside function. The correlation function $C(x)=2/\pi\sin(\pi x/2\xi)/(x/\xi)^{2}$
decays in an oscillatory fashion (inset). The potential $V(x)$ varies
smoothly around zero. \textbf{B:} A potential with exponential power
spectrum $S(k)=2\xi\exp\left[-2\xi|k|/\pi\right]$ and Lorentzian
correlation function $C(x)=\left[1+(\pi x/2\xi)^{2}\right]^{-1}$.
The potential $V(x)$ shows more structure on a finer scale. \textbf{C:}
Copolymers with different relative concentrations $c$ of monomer
types (gray and black). }
\end{figure}
In the following we investigate the relaxation properties of Eq.~(\ref{eq:FFPE})
in random potentials $V$. Since the shape of the potential is determined
by the ordering of different types of monomers along the chain, $V(x)$
will be bounded and fluctuate about some average. Furthermore, it
will generally possess a typical correlation length $\xi$. Without
loss of generality we let $\left\langle V\right\rangle =0$ and $\left\langle V^{2}\right\rangle =V_{0}^{2}$.
The correlation length $\xi$ is defined by $\xi=V_{0}^{-2}\int_{0}^{\infty}\dy\, C(y)$
where $C(y)=\left\langle V(y)V(0)\right\rangle $ is the correlation
function. The most straightforward way to incorporate these attributes
into a model is by using Gaussian random phase potentials,\begin{equation}
V(x)=\frac{1}{2\pi}\int\dk\,\phi(k)\, e^{-i(kx+\vartheta(k))},\label{eq:rphdef}\end{equation}
which are defined by a set of random uncorrelated phases $\vartheta(k)$
(with $\vartheta(k)=-\vartheta(-k)$) and the power spectrum $S(k)=\int\dx\, e^{ikx}C(x)$
(with $\phi(k)\phi^{\star}(k^{\prime})=2\pi S(k)\delta(k-k^{\prime})$).
The pdf associated with this choice of $V(x)$ is Gaussian with zero
mean and variance $V_{0}^{2}=(2\pi)^{-1}\int\dk\, S(k)$. Fig.~\ref{cap:fig2}A.
and~\ref{cap:fig2}B. show two realizations of random phase potentials,
each one with a different power-spectrum (and correlation function).

The relaxation properties are determined by the eigenvalue spectrum
$E(k)$ of the evolution operator $\op{L}$ defined by the rhs of
Eq.~(\ref{eq:FFPE}). In order to compute the spectrum, the FFPE
can be transformed by means of $p(x,t)=e^{-\beta V(x)/2}\psi(x,t)$.
This yields a fractional Schrödinger equation with identical spectral
properties,\begin{eqnarray}
\pdt\psi & = & -\op{H}\psi\qquad\textrm{with}\label{eq:FSE}\\
\op{H} & = & D\left(-\fraclpmu{}+U\right).\label{eq:FH}\end{eqnarray}
The operator $\op{H}$ is symmetric and the effective potential $U$
is related to the original potential $V$ by $U(x)=e^{\epsilon v}\fraclpmu{}e^{-\epsilon v}$,
where $v(x)=V/V_{0}$ is a rescaled potential of unit variance and
$\epsilon=\beta V_{0}/2=V_{0}/2k_{B}T$ is the potential strength
in units of $k_{B}T$. 

For vanishing potential $\epsilon=0$, we have $U\equiv0$ and $\op H_{0}=D\fraclpmu{}$
which describes free superdiffusion when $\mu<2$. The spectrum of
$\op H_{0}$ is given by $E_{0}(k)=Dk^{\mu}$. The wave number $k>0$
defines the spatial scale of the corresponding mode. When a potential
is present, the spectrum can be written as $E(k)=D_{\mu}(k;\epsilon)\, k^{\mu}$
where $D_{\mu}(k;\epsilon)$ quantifies the relaxation properties
on scales $\simeq k^{-1}$ with the unperturbed $k^{\mu}$-behavior
as a reference. If $D_{\mu}(k;\epsilon)/D<1$ the process relaxes
more slowly compared to free superdiffusion. The spectrum $E(k)$
can be obtained for weak potentials by perturbation theory. If $\epsilon\ll1$,
the effective potential $U$ can be treated as a small perturbation,
for $U=\mathcal{O}(\epsilon)$. Up to second order in $\epsilon$
the quantity $D_{\mu}(k;\epsilon)$ reads\begin{equation}
D_{\mu}(k;\epsilon)/D=\left(1-4\epsilon^{2}G_{\mu}(k)\right),\label{eq:gendefk}\end{equation}
where the effect on relaxation is provided by the function\begin{align}
G_{\mu}(k) & =\frac{1}{8\pi}\int\diff{q}\, S(q)\, g_{\mu}(k/q)\quad\textrm{with}\label{eq:bigGdef}\\
g_{\mu}(z) & =\frac{1}{z^{\mu}}\left(\frac{1}{(1+z)^{\mu}-z^{\mu}}+\frac{1}{|1-z|^{\mu}-z^{\mu}}-2\right).\label{eq:smallgdef}\end{align}
\begin{figure}
\includegraphics[%
  width=1.0\columnwidth]{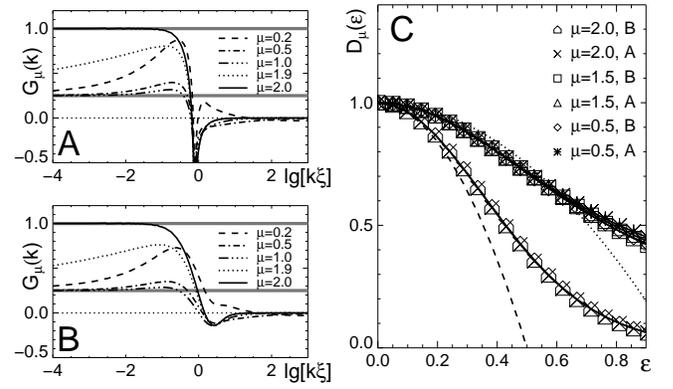}

\caption{\label{cap:fig3} Ralaxation for various Lévy exponents $\mu$(inset)
and random phase potentials. Panel \textbf{A} (\textbf{B}) corresponds
to the potential in figure~\ref{cap:fig2}A(B). $\lg[k\xi]$ denotes
the decadic logarithm. The gray lines indicate the limits $G_{\mu}(k\rightarrow0)$
for $\mu=2$ (upper line) and $\mu<2$ (lower line), see Eq.~(\ref{eq:limG}).
Panel \textbf{C} depicts the generalized diffusion coefficient $D_{\mu}(\epsilon)$
for the potential in figure~\ref{cap:fig2}A(B) and for three Lévy
exponents $\mu$. The dashed and dotted lines are the results obtained
from perturbation theory, i.e. $D_{\mu}(\epsilon)=1-4\epsilon^{2}$
($\mu=2$) and $D_{\mu}(\epsilon)=1-\epsilon^{2}$ ($\mu<2$). }
\end{figure}
Figs.~\ref{cap:fig3}A and \ref{cap:fig3}B depict $G_{\mu}(k)$
as a function of $k$ in units of the inverse correlation length $\xi^{-1}$
for the two types of random phase potentials defined in Fig.~\ref{cap:fig2}A
and \ref{cap:fig2}B. The solid line depicts the limiting case of
ordinary diffusion ($\mu=2$). The potential slows down the ordinary
diffusion process ($G_{\mu}(k)>0$) on scales larger than the correlation
length, and speeds it up ($G_{\mu}(k)<0$) on scales smaller than
$\xi$. The function $G_{\mu}(k)$ has a pronounced minimum at $k\approx\xi^{-1}$.
Moderately superdiffusive processes ($\mu\gtrsim1$) behave in a similar
fashion, exhibiting the highest variation for $k\approx\xi^{-1}$.
On the other hand, $G_{\mu}(k)$ differs strongly for different $\mu$
in the asymptotic regime $k\ll\xi^{-1}$. Note also that on small
scales ($k>\xi^{-1}$) almost all processes relax faster than without
the potential. In fact, $G_{\mu}(k)<0$ for $2\ge\mu\ge\mu_{c}$ where
$\mu_{c}=2-\ln3/\ln2\approx0.415$. Surprisingly, this is no longer
valid for strongly superdiffusive processes with $\mu<\mu_{c}$. For
instance, in the case $\mu=0.2$ (dashed line in Figs.~\ref{cap:fig3}A
and~\ref{cap:fig3}B) $G_{\mu}(k)$ is positive for $k>\xi^{-1}$,
implying that strongly superdiffusive processes are slowed down even
on small scales. Comparing potential types, we see that the relaxation
is different for each potential, but these differences become less
important in the asymptotic regime, which is governed by $G_{\mu}(k)$
as $k\rightarrow0$. This limit can be computed from Eqs.~(\ref{eq:bigGdef},
\ref{eq:smallgdef}), observing that $\int_{0}^{\infty}\diff{q}\, S(q)=\pi$
and $g_{2}(z\rightarrow0)=1$ and $g_{\mu<2}(z\rightarrow0)=1/4$,\begin{equation}
\lim_{k\rightarrow0}G_{\mu}(k)=\begin{cases}
1/4 & \quad\mu<2\\
1 & \quad\mu=2,\end{cases}\label{eq:limG}\end{equation}
Hence, the asymptotic behavior is universal, with the exception of
the limiting case of ordinary diffusion, and is independent of properties
of the potential. The range of validity of the limit~(\ref{eq:limG}),
however, strongly depends on $\mu$. The limit is not attained for
marginal exponents (e.g. $\mu=0.2$ and $1.9$) even on scales several
orders of magnitude larger than the correlation length (Fig.~\ref{cap:fig3}A
and~\ref{cap:fig3}B). The above results are valid for small potential
strengths $\epsilon$. For higher effective potential strengths we
investigate the asymptotics numerically. The quantity of interest
is the normalized generalized diffusion coefficient $D_{\mu}(\epsilon)$
defined by\begin{equation}
D_{\mu}(\epsilon)=\lim_{k\rightarrow0}D_{\mu}(k;\epsilon)/D.\label{eq:genDdef}\end{equation}
In the perturbative regime Eqs.~(\ref{eq:gendefk}) and~(\ref{eq:genDdef})
yield the universal relation $D_{\mu}(\epsilon)=1-\epsilon^{2}$ for
$\mu<2$ and $D_{\mu}(\epsilon)=1-4\epsilon^{2}$ for ordinary diffusion.
Figs.~\ref{cap:fig3}C and~\ref{cap:fig3}D compare these results
to those obtained numerically. Although the numerics deviates from
perturbation theoretic predictions for greater potential strengths
$\epsilon$, the universality still holds, i.e. the asymptotics ($k\rightarrow0$)
is independent of $\mu$ and of the statistical properties of the
potential. The crucial property is the non-locality of the process
(i.e. $\mu=2$ vs. $\mu\ne2$). Thus, as soon as the folding properties
of the chain permit scale free transitions ($\mu\neq2$), the behavior
of $D_{\mu}(\epsilon)$ changes abruptly.

\begin{figure}
\begin{center}\includegraphics[%
  width=1.0\columnwidth]{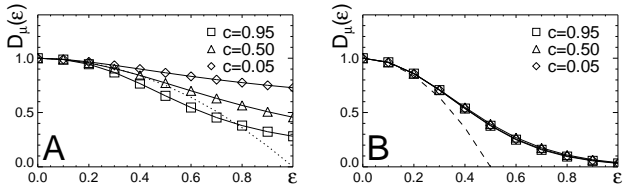}\end{center}

\caption{\label{cap:fig4} The generalized diffusion coefficient $D_{\mu}(\epsilon)$
for the copolymer potential, i.e. Eq.~(\ref{eq:copopot}) at three
relative monomer concentrations. The potentials (Fig.~\ref{cap:fig2}C)
either possess sparsely distributed peaks ($c=0.05$), troughs ($c=0.95$)
or vary uniformly ($c=0.5$). Dashed ($\mu=2$) and dotted ($\mu=1$)
lines represent perturbation theoretic results. \textbf{A:} $D_{\mu}(\epsilon)$
for a Lévy flight ($\mu=1$) is different for each potential. \textbf{B:}
$D_{\mu}(\epsilon)$ for ordinary diffusion ($\mu=2$) is independent
of $c$, the curves coincide.}
\end{figure}
The pdf of random phase potentials is symmetric with respect to the
mean, i.e. a value $V$ is as likely to occur as $-V$ along the chain.
For a number of hetero-polymers this assumption is inadequate. Consider
the simple model copolymer depicted in Fig.~\ref{cap:fig2}C. The
chain consists of a random arrangement of monomers, each one equipped
with an intrinsic local potential parity $v_{-}$ and $v_{+}$, with
$-v_{-}=v_{+}>0$ and an interaction range which we assume to be a
Gaussian $f(x-x_{n})$ centered at the monomer site $x_{n}$,\begin{equation}
V(x)=\sum v_{n}\, f(x-x_{n}),\label{eq:copopot}\end{equation}
with $f(x)=\exp[-x^{2}/2\sigma^{2}]$ and $\sigma\approx a$. The
$v_{n}$ are randomly drawn from a pdf $p(v)=c\delta(v-v_{-})+(1-c)\delta(v-v_{+})$.
The relative concentration of low and high energy monomers is given
by $c$ and $1-c$, respectively. The parameter $c$ determines the
shape of the overall potential. When $c<1/2$ ($c>1/2$) the potential
consists of a series of localized peaks (troughs). Mean and variance
of the potential are $\left\langle V\right\rangle =(1-c)\, v_{-}+c\, v_{+}$
and $\left\langle (V-\left\langle V\right\rangle )^{2}\right\rangle =(1-c)\, c\,\delta v^{2}$
with $\delta v=v_{+}-v_{-}$. Figs.~\ref{cap:fig4}A and~\ref{cap:fig4}B
depict the results obtained for the generalized diffusion coefficient
$D_{\mu}(\epsilon)$ on these types of copolymers for three values
of $c$, each one representing one of the situations depicted in Fig.~\ref{cap:fig2}C.
The parameter $\delta v$ was chosen such that the variance is identical
in all potentials. Although the value of $D_{\mu}(\epsilon)$ in the
weak potential regime ($\epsilon\ll1$) is consistent with the one
observed in random phase potentials, for greater values of $\epsilon$
a striking deviation occurs. On one hand, the ordinary diffusion process
($\mu=2$) is nearly insensitive to the shape of the potential, all
functions $D_{2}(\epsilon)$ coincide. On the other hand, the superdiffusive
process exhibits a more (less) pronounced decrease with increasing
$\epsilon$ when $c=0.95$ ($c=0.05$) as compared to an unbiased
concentration of monomer types. 

The results reported in this Letter predict for superdiffusive behavior
on folding polymers that the dispersion speed depends strongly on
the specific arrangement of various types of monomers. This is in
sharp contrast to the case of ordinary diffusion, which solely depends
on magnitude variations of the potential. Therefore we expect that
potential heterogeneity is an essential ingredient in superdiffusive
translocation phenomena of proteins along biopolymers.

\begin{acknowledgments}
D. B. thanks W. Noyes for interesting comments and discussion.
\end{acknowledgments}
\bibliographystyle{apsrev}
\bibliography{/home/zwerg/bibliography/bib/books,/home/zwerg/bibliography/bib/paper}

\end{document}